\documentclass[aps,pra,showpacs,twocolumn,groupedaddress]{revtex4}
\usepackage{graphicx}
\usepackage{amsmath}
\usepackage{epstopdf}
\usepackage{amsfonts}
\usepackage{amssymb}

\begin{document}

\title{Beyond the Dicke Quantum Phase Transition with a Bose-Einstein Condensate in an
Optical Ring Resonator}
\author{D. Schmidt, H. Tomczyk, S. Slama, C. Zimmermann}
\affiliation{Physikalisches Institut, Eberhard-Karls-Universit\"{a}t T\"{u}bingen, Auf der
Morgenstelle 14, D-72076 T\"{u}bingen, Germany.}
\begin{abstract}
We experimentally investigate the dynamical instability of a Bose Einstein
condensate in an optical ring resonator for various cavity detuning and pump
powers. The resulting phase diagram is asymmetric with respect to the cavity
detuning and can be described by the coupling of two atomic modes with one
optical mode. We compare the experimental data to a numerical simulation and
to an analytic expression of the phase boundary. For positive and negative
pump cavity detuning different coupling mechanisms are identified explaining
the asymmetry of the phase diagram. We present a physical interpretation and
discuss the connection to the Dicke quantum phase transition.
\end{abstract}
\maketitle

A cloud of cold atoms interacting with the light field inside an optical resonator has turned out to be a surprisingly rich test bed for investigating fundamental concepts
such as collective light scattering, self organization, superradiance, optomechanics, and cavity cooling \cite{Ritsch10}. Recently, the field has gained extra attention by the experimental realization of the Dicke quantum phase transition. Originally, the Dicke model was developed to describe a collection of two level atoms interacting with a single optical mode. It predicts a phase transition from a normal to a superradiant phase at finite temperatures which has a quantum analog also in the limit of zero temperature \cite{Hepp73}\cite{Wang73}. The critical coupling for the phase transition is hard to reach with optical single photon transitions, however with cavity enhanced Raman coupling the threshold is drastically reduced \cite{Dimer07} and the experimental observation becomes feasible. In fact, a Dicke phase transition has recently been demonstrated with a Bose-Einstein condensates in an optical standing wave resonator \cite{Zurich}\cite{Mottl12}\cite{Nagy10}\cite{Dom02}. A similar dynamical instability was also observed some time ago with a condensate in a longitudinally pumped optical ring cavity \cite{Slama07} and was interpreted as a quantum version of collective atomic recoil lasing. This kind of laser forms the atomic analog to the free electron laser and has been studied in great detail since more than a decade \cite{Piovella01}\cite{Ritsch10}.

In this paper, we experimentally investigate the threshold of atomic recoil lasing and show that it's physical origin is very similar to that of the Dicke phase transition. Both instabilities can be traced back to a term in the interaction Hamiltonian which describes the simultaneous creation of a photon in the optical mode and an atom in a state of finite momentum \cite{Emary03b}. In the case of collective recoil lasing, the optical light mode couples to two atomic states with opposite momentum. One obtains a three mode scenario \cite{Moore99} with aspects that go beyond the standard quantum Dicke model. Depending on the detuning $\Delta=\omega_{c}-\omega$ between the pump laser frequency $\omega$ and the resonance frequency of the resonator $\omega_{c}$, two different physical situations arise: For negative detuning $\Delta$, a photon is created together with an excitation of an atom into one of the finite momentum states and atomic recoil lasing is observed \cite{Slama07}: At threshold, the atoms scatter photons from the pump cavity mode into the backwards propagating probe mode and arrange inside an emerging moving dipole lattice potential. In turn, the resulting density grating
enhances the scattering efficiency which gives rise to a dynamical instability. For positive detuning $\Delta$ the situation is very different since now the two atomic states are simultaneously excited. In this regime, the optical mode can be adiabatically eliminated and the phase transition is caused by light induced correlation between atoms of opposite momenta. At threshold, they form a stationary density grating with the position of the maxima subject to spontaneous symmetry breaking. Furthermore, the coupling is
accompanied by mode softening of the momentum states. The existence of a recoil lasing regime and a momentum correlated regime leads to a phase diagram which is strongly asymmetric with respect to the pump cavity detuning. This asymmetry is the feature that we observe experimentally.

\begin{figure}[ht]
\label{LaserSystemCARL} 
\includegraphics[width=\columnwidth]{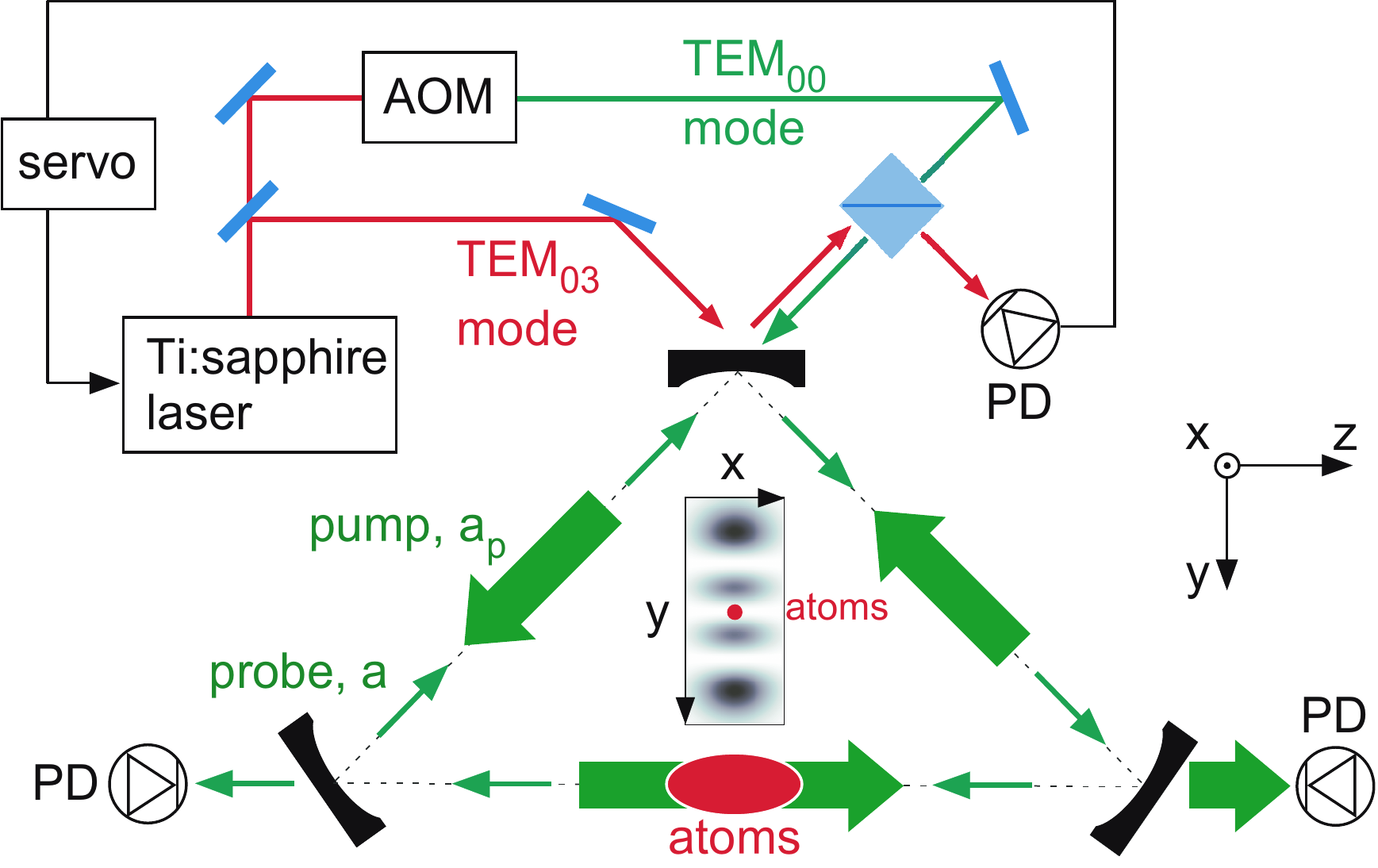}
\caption{(color online) Experimental setup. A Bose-Einstein condensate is placed in an optical ring resonator and exposed to the light field of a low finesse TEM$_{00}$ pump mode. The pump light is provided by a Ti:sapphire laser which is locked via Pound-Drever-Hall technique to a TEM$_{03}$ high finesse reference mode. Its beam profile with a minimum at the position of the condensate (inset) has a negligible effect on the atoms. With an acousto-optic modulator (AOM) the light is shifted by 130 $\operatorname{MHz}$ and tuned across the resonance of the low finesse TEM$_{00}$ mode. The AOM also controls the power of the pump light. The atoms scatter light from the pump mode into the reverse propagating probe mode. Two photodiodes monitor the light that leaks out through the high reflecting mirrors.}
\label{setup}
\end{figure}

The experiment is sketched in figure \ref{setup}. A $^{87}$Rb Bose-Einstein condensate is captured in a Ioffe-type magnetic trap at the focus of a horizontally oriented ring cavity \cite{Slama07}. The volume of the TEM$_{00}$ mode is given by $V=\frac{1}{2}\pi Lw_{x}w_{y}=1.37$ mm$^{3}$ with beam radii of $w_{x}=117\operatorname{\mu m}$ (vertical) and $w_{y}=88\operatorname{\mu m}$ (horizontal). The round trip length of $L=87\operatorname{mm}$ leads to a free spectral range of $\nu_{fsr}=3.45\operatorname{GHz}$. The finesse of the cavity amounts to $130000$ (high finesse) for s polarized light and $2800$ (low finesse) for p polarized light. The cavity is longitudinally pumped by a Ti:sapphire laser on the low finesse TEM$_{00}$ mode (field decay rate $\kappa=2\pi\times 650 \operatorname{kHz}$) which is red detuned relative to the rubidium D1 line ($F=2\rightarrow F^{\prime}=1,2$) by $91.6\operatorname{GHz}$. 

After the condensate has been prepared with an atom number of about $N=8\times10^{4}$, the pump light is switched on for $50\operatorname{\mu s}$. Then, the atoms are released from the trap and an absorption image is taken after $9\operatorname{ms}$ of ballistic expansion. For pump powers $P_{p}$ above a critical value, the dynamical instability occurs and a macroscopic number of atoms scatter photons from the pump mode into the probe mode. The photon recoil kicks the atoms from the condensate into a state with a finite momentum. 

\begin{figure}[th]
\includegraphics[width=\columnwidth]{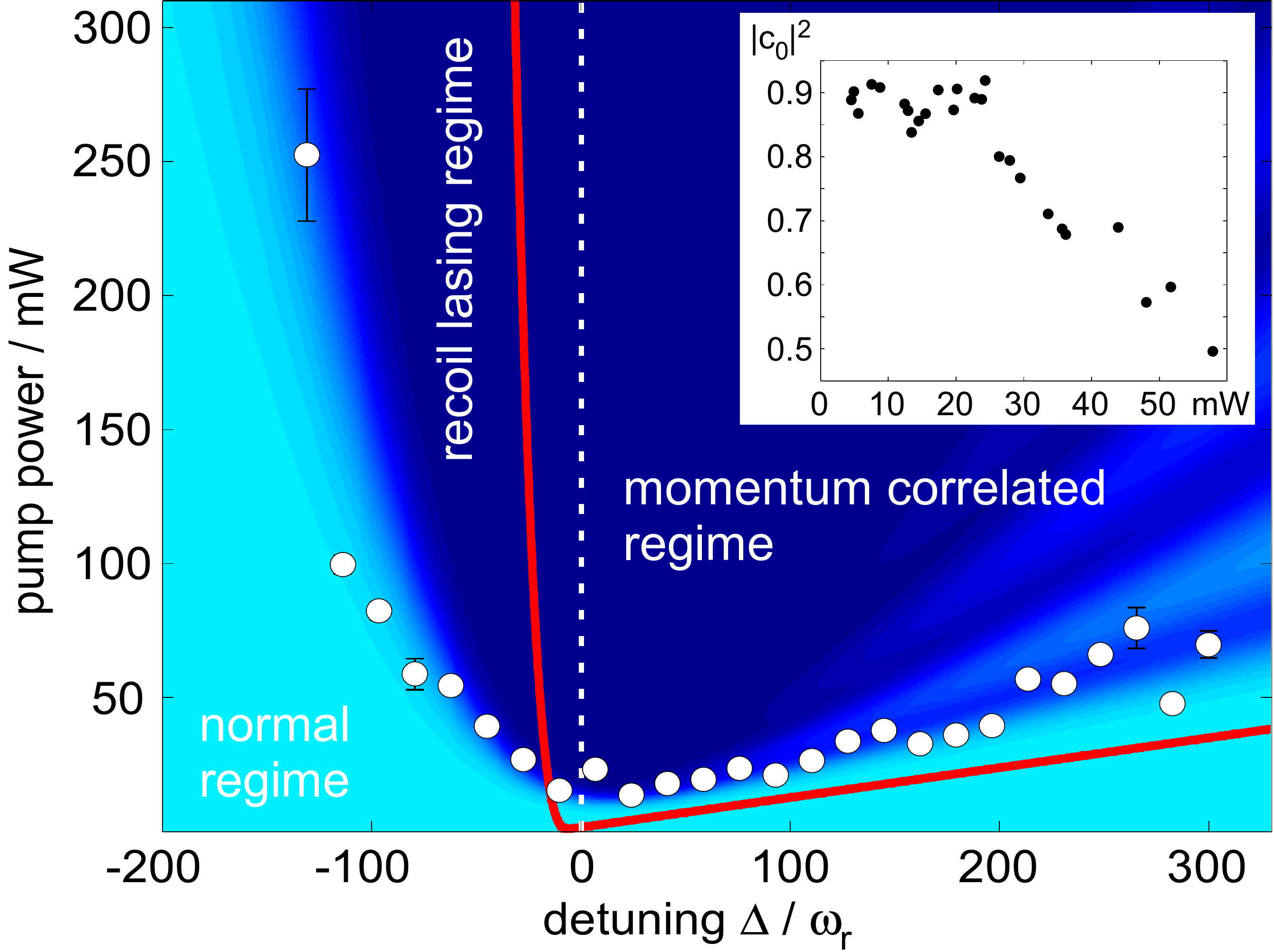}
\caption{(color online) Phase diagram. The critical circulating power in the pump mode which is needed to deplete the condensate population $\left|c_{0}\right|^{2}$ by $(30\pm10)$\% is plotted for various pump cavity detuning (white dots). The color code shows the results of the numerical simulation of equation \eqref{allgemeineBewegungsgleichungen}, pale blue: $\left|c_{0}\right|^{2}=1$, dark blue $\left|c_{0}\right|^{2}=0$. The red line marks the phase boundary as determined from the three mode model without cavity damping. The dashed line separates the different coupling regimes. Inset: variation of condensate population with pump power for $\Delta =  100\operatorname{\omega_r}$.}
\label{phasendiagramm}
\end{figure}

The scattered atoms are identified in the absorption images and their relative fraction is determined. For each value of the detuning $\Delta$, we vary the power in the pump mode until the condendsate population is depleted by $(30\pm10)$\%. The such determined critical pump power is plotted in fig. \ref{phasendiagramm}. For each detuning, an average of three measurements is taken. The error bars indicate the standard deviation.

For the theoretical description of the experiment we combine the analysis of \cite{Moore99}, \cite{Dimer07}, and \cite{Piovella01}. The photons in the pump mode are represented by the field operator $\hat{A}_{p}=\hat{a}_{p}e^{ikz}$ and the photons in the reverse mode by $\hat{A}=\hat{a}e^{-ikz}$. The atomic matter field $\hat{\psi}=\sum\hat{c}_{n}e^{2inkz}$ is expanded into momentum eigenstates separated by $2\hbar k$ which is the momentum transferred to the atoms by scattering a photon. $\hat{\psi}$ is normalized to the total atom number $N$. By using the plane waves basis, finite size effects are neglected for the atoms and for the light. The interaction between the atoms and the light is given by the optical dipole potential which in second quantization reads 
$H_{int}=U_{0}\int \hat{\psi}^{+}\hat{\psi}\left(\hat{A}_{p}+\hat{A}\right)\left(\hat{A}_{p}^{+}+\hat{A}^{+}\right)dV$. The interaction strength is given by the single photon light shift $U_{0}=\frac{1}{\Delta_{a}}\frac{\omega d^{2}}{2\varepsilon_{0}V}$ with the dipole moment of the atomic transition $d$, the cavity mode volume $V$, the dielectric constant $\varepsilon_{0}$, the frequency of the pump beam $\omega$, and the detuning of the pump beam $\Delta_{a}=\omega-\omega_{0}$ relative to the resonance frequency of the atomic transition $\omega_{0}$. 

The equations of motion are derived from the Hamilton $H=H_{0}+H_{int}$ with $H_{0}=\int\left(\psi^{+}\left(-\frac{\hbar^{2}\nabla^{2}}{2m}\right)\psi+\hbar\Delta\left(A^{+}A+A_{p}^{+}A_{p}\right)\right)dV$. The contribution from atomic collisions \cite{Moore99} is small and can be neglected. In the mean field approximation the operators are replaced by their expectation values $a_{p}$, $a$, and $c_{n}$. We further assume a constant optical pump mode and obtain
\begin{align}
\dot{c}_{n}  &  =-in^{2}\omega_{r}c_{n}-i\frac{U_{0}}{\hbar}\left(
c_{n-1}a_{p}a^{\ast}+c_{n+1}aa_{p}^{\ast}\right)
\label{allgemeineBewegungsgleichungen}\\
\dot{a}  &  =\left(  -i\Delta-i\frac{U_{0}}{\hbar}N\right)  \cdot
a-i\frac{U_{0}}{\hbar}a_{p}\sum_{n}c_{n}^{\ast}c_{n-1}-\kappa a\nonumber
\end{align}
with the atomic recoil frequency $\omega_{r}=\frac{1}{\hbar}\frac{\left(2\hbar k\right)  ^{2}}{2m}=2\pi\cdot14.5\operatorname{kHz}$ and the total number of atoms 
$N=\sum c_{n}^{\ast}c_{n}$. The finite cavity linewidth is taken into account by adding the decay term $-\kappa a$. Light that is scattered into the probe mode interferes with the pump light and generates an optical lattice potential $U_{dip}(z)=U_{0}\left|  a_{p}e^{ikz}+ae^{-ikz}\right|  ^{2}$ which can be stationary or moving depending on the time dependence of $a$. Similarly the interference between the atomic momentum states generates a periodic density grating given by 
$\psi\psi^{\ast}=\sum_{n,m}c_{n}^{\ast}c_{m}e^{2i\left(  m-n\right)  kz}$. At threshold almost all atoms occupy the condensate state $c_{0}$ and one can approximate the sum in equation \eqref{allgemeineBewegungsgleichungen} by keeping only terms which contain $c_{0}$ or $c_{0}^{\ast}$, i.e. $n=0,\pm1$. Once the instability has started, transitions into higher momentum states are possible, however, they do not contribute to the threshold behaviour.

The four remaining equations are solved numerically with $N=8\times10^{4}$, $U_{0}=\hbar\cdot2\pi\cdot 0.10\operatorname{Hz}$, and $\kappa=45 \operatorname{\omega_{r}}$ (fig. \ref{phasendiagramm}). For a given detuning $\Delta$ and increasing circulating power in the pump mode $P_{p}=\left|  a_{p}\right|  ^{2}\hbar\omega\nu_{fsr}$, there is a narrow range at which the population of the condensate drops from 1 (pale blue) to 0 (dark blue) and the system undergoes a non equilibrium phase transition from the normal to a collectively excited state. The boundary between these two regimes is well reproduced by the experimental data (white dots). The simulations critically depend on the initial photon number $\left|  a\right|  ^{2}$ in the probe mode and the initial atom number $\left|  c_{\pm1}\right|  ^{2}$of the two states with finite momentum. Best agreement with the data is obtained for 
$\left|a\right|^{2}=1$ and $\left|c_{1}\right|^{2}=$ $\left|c_{-1}\right|^{2}=5$. This is consistent with a thermal occupation of the two atomic states and quantum fluctuations in the optical mode. Other possible triggers for the instability are technical noise and Mie scattering due to the finite size of the condensate \cite{Bach12}. 

The connection to the Dicke model can be made by assuming that at threshold the momentum state with $n=0$ (condensate) and the pump mode are macroscopically occupied and constant in time. The operators $\hat{c}_{0}$ and $\hat{a}_{p}$ can then be replaced by their constant mean field values $c_{0}$ and $a_{p}$ and from $H_{int}$ one obtains an effective interaction Hamiltonian \cite{Moore99} which contains only the modes $\hat{a}$, $\hat{c}_{1}$, and $\hat{c}_{-1}$.
\begin{equation}
H_{3m}=\lambda\left(  \hat{a}\hat{c}_{1}+\hat{a}^{+}\hat{c}_{1}^{+}
+\hat{a}\hat{c}_{-1}^{+}+\hat{a}^{+}\hat{c}_{-1}\right)  \label{DickeHamilton}
\end{equation}
Here, the coupling strength is $\lambda=U_{0}\sqrt{N_{p}N_{0}}$ with the number of photons in the pump mode $N_{p}=\left|a_{p}\right|^{2}$ and the number of atoms in the initial condensate $N_{0}=\left|  c_{0}\right|  ^{2}$. The interaction Hamiltonian $H_{3m}$ can be compared to the Dicke Hamiltonian in the Holstein-Primakoff representation in the limit of large atom numbers \cite{Emary03a}. It differs from $H_{3m}$ only in the sign of the index in the last two terms. The first two terms appear also in the Dicke Hamiltonian and are responsible for the dynamical instability \cite{Emary03b}. In a closed system each of the two terms seem to violate energy since two quanta are created/annihilated simultaneously without a corresponding annihilation/creation. However, in our experiment the resonator is externally pumped and the interpretation is straight forward: By scattering a photon from the pump mode into the probe mode, an atom is scattered from the condensate into a state with $n=1$ and vice versa. The last two terms describe the energy exchange between the probe mode and the atomic state with $n=-1$ and also have a clear physical interpretation: A photon in the probe mode is destroyed and scattered back into the pump mode while an atom is kicked out of the condensate into a state with momentum $-2\hbar k$ and vice versa. The asymmetry of the phase diagram with respect to $\Delta$ is a direct consequence of including the atomic state with $n=-1$.

To better understand the underlying physics, we take a closer look at the equations of motion for the Hamiltonian $H=H_{0}+H_{3m}$ \cite{Moore99}
\begin{equation}
i\frac{d}{dt}\left(
\begin{array}
[c]{c}
c_{-1}\\
a\\
c_{1}^{\ast}
\end{array}
\right)  =\left(
\begin{array}
[c]{ccc}
\omega_{r} & \lambda & 0\\
\lambda & \Delta & \lambda\\
0 & -\lambda & -\omega_{r}
\end{array}
\right)  \left(
\begin{array}
[c]{c}
c_{-1}\\
a\\
c_{1}^{\ast}
\end{array}
\right)  .\label{Bewegungsgleichung}
\end{equation}
The finite cavity life time is neglected and the photons in the probe mode now have a well defined energy $\hbar\Delta$. The matrix couples $c_{-1}$ and $a$ with off diagonal elements $\lambda$ of the same sign. This corresponds to a two level system with conventional coupling. The energy levels thus repel each other with increasing coupling strength $\lambda$. In contrary, $a$ and $c_{1}^{\ast}$ are coupled with off diagonal elements $\lambda$ of opposite sign which can be mapped to a two level system with imaginary coupling strength $i\lambda$. In such a system the energy levels of the two states attract each other and eventually become degenerate at a critical coupling
strength. For even larger coupling the eigenvalues develop an imaginary component which gives rise to an exponential instability. This unconventional coupling between $a$ and $c_{1}^{\ast}$ is caused by the first two terms of $H_{3m}$.

\begin{figure}[ht]
\includegraphics[width=\columnwidth]{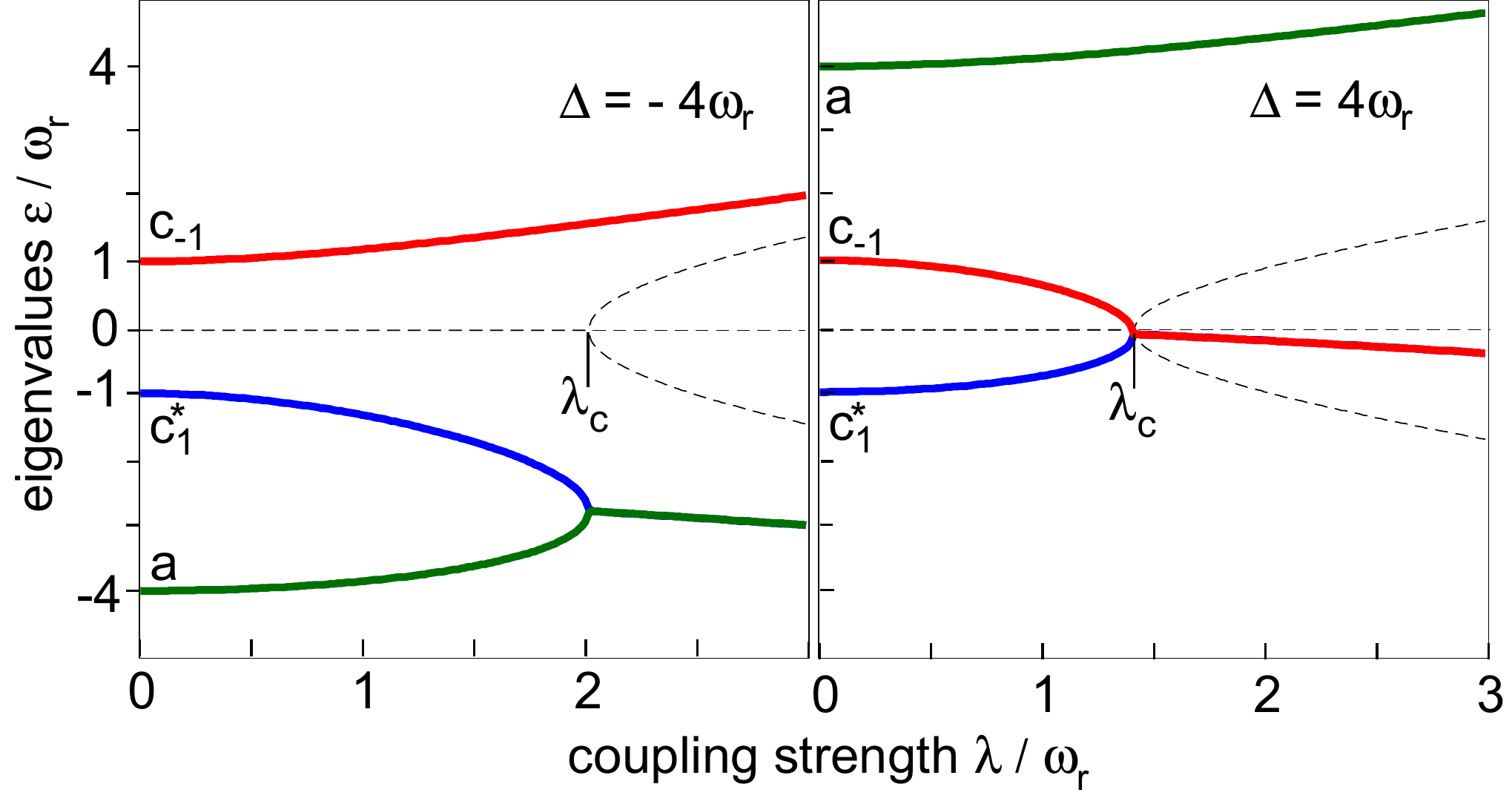}
\caption{(color online) Eigenfrequencies of the three mode model. At zero coupling, $\lambda=0$, the three modes have a purely optical or atomic character ($a$, $c_{1}^{\ast}$,
$c_{-1}$) with the corresponding eigenfrequencies $\Delta$ and $\pm\omega_{r}$. With increasing coupling strength the states are coupled. The two frequencies of the two low lying states merge and become identical at the critical coupling strength $\lambda_{c}$. For negative detuning $\Delta$ the optical mode merges with an atomic state (recoil lasing regime) while for positive detuning the two atomic states merge (momentum correlated regime). This leads to the strong asymmetry of the phase diagram. The dashed lines show the imaginary components of the eigenvalues which become nonzero at critical coupling.}
\label{eigenwerte}
\end{figure}

The Ansatz $\vec{\varphi}=\vec{\varphi}_{\varepsilon}e^{-i\varepsilon t}$ solves equation (\ref{Bewegungsgleichung}) with $\varepsilon$ and $\vec{\varphi}_{\varepsilon}$ being the eigenvalues and eigenvectors of the matrix. Figure \ref{eigenwerte} shows the spectra $\varepsilon\left(\lambda\right)  $ of the three eigenstates $\vec{\varphi}_{\varepsilon}$ for two different values of detuning. For $\lambda=0$, the eigenvalues are real and correspond to the frequencies of the purely optical state, $\vec{\varphi}_{\varepsilon=\Delta}=\left(0,a,0\right)  $, and two  purely atomic states, $\vec{\varphi}_{\varepsilon=\omega_{r}}=\left(  c_{-1},0,0\right)  $, and $\vec{\varphi}_{\varepsilon=-\omega_{r}}=\left(  0,0,c_{1}^{\ast}\right)  $. For a finite coupling, $\lambda>0$, the three states get mixed and loose their pure character. The frequencies of the two lowest states converge and eventually become identical
at the critical coupling strength
\begin{equation}
\tilde{\lambda}_{c}^{2}=\frac{1}{27}\left(  9\tilde{\Delta}-\tilde{\Delta}%
^{3}+\left(  3+\tilde{\Delta}^{2}\right)  ^{3/2}\right),
\label{Phasengrenze}
\end{equation}
with $\tilde{\Delta}=\Delta/\omega_{r}$ and $\tilde{\lambda}_{c}=\lambda_{c}/\omega_{r}$. At critical coupling also the two eigenstates are identical. For 
$\left| \lambda\right|  >\lambda_{c}$ the eigenvalues of the converging states become complex and one of them diverges exponentially in time while the
other decays. Equation \eqref{Phasengrenze} thus marks the boundary of the normal phase. (red line in fig.\ref{phasendiagramm}).

For positive cavity detuning, equation \eqref{Phasengrenze} seems to predict a lower threshold as compared to the observations and to the numerical simulations.
However, the threshold is observable in a finite time only if the exponential growth rate is sufficiently large. This shifts the experimental threshold slightly above the theoretical value. From figure \ref{eigenwerte} one can see that for $\Delta>\omega_{r}$ the two atomic states with $n=\pm1$ merge at a frequency $\varepsilon=0$. Above threshold they thus form a stationary density grating which grows from the initial fluctuations with an arbitrary position of the nodes. This behavior prevails also for a finite cavity decay rate: For $\kappa\gg\lambda^{2}/\left|  \Delta\right|  $ the equation for the optical mode can be adiabatically eliminated by setting $da/dt=0$ and, in the limit of 
$\Delta\gg\kappa$, the eigenvalues $\varepsilon^{2}=1-\frac{2\lambda^{2}}{\Delta^{2}+\kappa^{2}}\left(  i\kappa+\Delta\right)  $ become independent of $\kappa$. The threshold now reads $\lambda_{c}^{2}=\omega_{r}\Delta/2$ which establishes the limiting case of \eqref{Phasengrenze} for large positive detuning. In this regime an entangled pair of two atoms are simultanuously excited into states of opposite momentum. The photon mediates the interaction between the atoms, but its actual probability to be found in the cavity is very small. The emerging correlation between two states of opposite momentum is specific to a single sided pumped ring resonator. To our knowledge this mechanism has not been identified before. It may offer new ways to study quantum gases with long range interaction \cite{Mottl12}.

For $\Delta<\omega_{r}$ the predicted nonlinear increase of the theoretical threshold \eqref{Phasengrenze} is also found in the experiment. A quantitative
agreement cannot be expected since the finite cavity linewidth cannot resolve the sharp rise which is thus smeared out over about one $\kappa$. In this regime the phase transition occurs when the atomic state with $n=1$ merges with the optical mode (fig. \ref{eigenwerte}). The merged state has a finite frequency $\varepsilon$ and generates a running density wave which is synchronized with a running optical potential. In this recoil lasing regime, the energy separation of the two converging modes at $\lambda=0$ now changes with the detuning $\Delta$. With increasing detuning, also the critical coupling $\lambda_{c}$ is pushed out to larger values which leads to the sharp increase at this side of the phase diagram. In the recoil lasing regime, adiabatic elimination of the optical mode is not a good approximation near the threshold. Since for large negative detuning the critical coupling strength grows as $\lambda_{c}^{2}\simeq2/27\cdot\left|  \Delta\right|  ^{3}/\omega_{r}$, the condition for adiabaticity is valid only for small detuning $\left|  \Delta\right|  \ll\sqrt{27/2\cdot\omega_{r}\kappa}$.

In summary, we have experimentally mapped out the phase diagram of a Bose Einstein condensate in a single sided pumped optical ring resonator. The diagram is interpreted with a three mode model that includes one optical mode and two atomic states with opposite momentum. For positive and negative cavity detuning, two different coupling mechanisms are identified and experimentally confirmed by the pronounced asymmetry of the phase diagram. For large positive detuning the optical mode can be adiabatically eliminated and the dynamics is dominated by the simultaneous excitation of entangled atom pairs in states of opposite momentum. In the future, high finesse cavities can be used to suppress transitions into higher momentum states ($n>1$) \cite{Wolke12}. In such closed systems the generation of novel stable quantum phases should be possible. In particular, it will be interesting to investigate the transition between the recoil lasing and the momentum correlated regime for $\lambda>\lambda_{c}$. 

We thank Andreas Hemmerich, Nicola Piovella, Daniel Braun and Nils Schopohl for stimulating discussions. This work has been supported by the Deutsche Forschungsgemeinschaft and the Research Executive Agency (program COSCALI, No. PIRSES GA-2010-268717). H.T. acknowledges support by the Evangelisches Studienwerk in Villigst.

\end{document}